\documentclass{PoS}
\usepackage{amsmath}
\usepackage{graphicx}
\usepackage[utf8]{inputenc}
\usepackage{subfigure}

\newcommand{\mpcac}{\ensuremath{m_{\text{\tiny PCAC}}}}
\newcommand{\arxiv}[2]{[arXiv:\,\href{http://arxiv.org/abs/#1}{\texttt{#1}} [\texttt{#2}]]}
\newcommand{\arxivold}[1]{[arXiv:\,\href{http://arxiv.org/abs/#1}{\texttt{#1}}\,]}

\title{Indications for infrared conformal behaviour of SU(2) gauge theory with $N_f = 3/2$ flavours of adjoint fermions}

\ShortTitle{Indications for infrared conformal behaviour of SU(2) gauge theory with $N_f = 3/2$ flavours of adjoint fermions}

\author{Georg Bergner\\
       University of Jena, Institute for Theoretical Physics, Jena, Germany
       }
\author{Pietro Giudice\\
       University of Münster, Institute for Theoretical Physics, Münster, Germany
       }
\author{Istvan Montvay\\
       Deutsches Elektronen-Synchrotron DESY, Hamburg, Germany
       }
\author{Gernot Münster\\
       University of Münster, Institute for Theoretical Physics, Münster, Germany
       }
\author{Stefano Piemonte\\
       University of Regensburg, Institute for Theoretical Physics, Regensburg, Germany
       }
\author{\speaker{Philipp Scior}\\
        University of Bielefeld, Fakultät für Physik, Bielefeld, Germany\\
        University of Münster, Institute for Theoretical Physics, Münster, Germany\\
        E-mail: \email{scior@physik.uni-bielefeld.de}}

\abstract{We present the results of a numerical investigation of SU(2) gauge theory with $N_f=3/2$ flavours of fermions, corresponding to 3 Majorana fermions, which transform in the adjoint representation of the gauge group. At two values of the gauge coupling, the masses of bound states are considered as a function of the PCAC quark mass. The scaling of bound states masses indicates an infrared conformal behaviour of the theory. We obtain estimates for the fixed-point value of the mass anomalous dimension $\gamma^*$ from the scaling of masses and from the scaling of the mode number of the Wilson-Dirac operator.}

\FullConference{The 36th Annual International Symposium on Lattice Field Theory - LATTICE2018\\
		22-28 July, 2018\\
		Michigan State University, East Lansing, Michigan, USA.}

\begin{document}

\section{Introduction}
The long and short distance behaviour of QCD-like theories depends significantly on the number $N_f$ of fermion flavours and on the representation of the gauge group under which the fermions transform. For sufficiently small $N_f$ the $\beta$-function is negative and the well-known scenario with confinement and asymptotic freedom occurs. If $N_f$ exceeds a certain limit $N_f^l$ the $\beta$-function shows asymptotic freedom near the origin but develops an infrared fixed point, called Banks-Zaks fixed point \cite{Banks:1981nn}, at a finite value of the coupling. If one further increases the number of flavours the IR fixed point moves to weaker coupling. Until, at a certain limit $N_f^u$ asymptotic freedom is lost, the $\beta$-function is positive and we find an IR fixed point at the origin. The region between $N_f^u$ and $N_f^l$, where the theory shows a conformal behaviour is called the conformal window. Its upper edge can be estimated with perturbation theory though the determination of the lower edge is a non-perturbative problem. \newline
It is the purpose of this article to present results about SU(2) gauge theory with $N_f = 3/2$ flavours of fermions in the adjoint representation of the gauge group, where $3/2$ means 3 flavours of Majorana (or Weyl) fermions. These results have also been presented in \cite{Bergner:2017gzw,Bergner:2017ytp}. We have investigated the masses of various particles, including mesons, glueballs and spin 1/2 fermion-glue bound states, the string tension, and the mass anomalous dimension, as well as the running coupling, in order to gain insights into the IR behaviour of the theory.
\section{Lattice Setup and Simulations}
We consider SU(2) gauge theory coupled to fermions transforming under the adjoint representation of the gauge group. The lattice formulation of the theory that we use employs the tree-level Symanzik improved gauge action and the Wilson-Dirac operator in the adjoint representation. The lattice action is
\begin{equation}
S_L =
S_G
+\sum_{xy,f} \bar{\psi}_x^{f}(D_w)_{xy}\psi_y^{f}\,,
\end{equation}
where $S_G$ is the gauge action and $D_w$ is the Wilson-Dirac operator
\begin{multline}
(D_w)_{x,a,\alpha;y,b,\beta}
    =\delta_{xy}\delta_{a,b}\delta_{\alpha,\beta}\\
    -\kappa\sum_{\mu=1}^{4}
      \left[(1-\gamma_\mu)_{\alpha,\beta}(V_\mu(x))_{ab}
                          \delta_{x+\mu,y}
      +(1+\gamma_\mu)_{\alpha,\beta}(V^\dag_\mu(x-\mu))_{ab}
                          \delta_{x-\mu,y}\right].
\end{multline}
Here the hopping parameter $\kappa$ is related to the bare fermion mass via $\kappa=1/(2m_0+8)$. The gauge field variables $V_{\mu}(x)$ in the adjoint representation are given by $[V_\mu(x)]^{ab}=2\,\mathrm{tr} [U^\dag_\mu(x) T^a U_\mu(x) T^b]$. For our simulations we use Majorana fermions, satisfying the Majorana condition
\begin{equation}
\overline{\psi} = \psi^{T} C,
\end{equation}
\begin{figure}[ht!]
\centering
\subfigure{\includegraphics[width=0.49\textwidth]{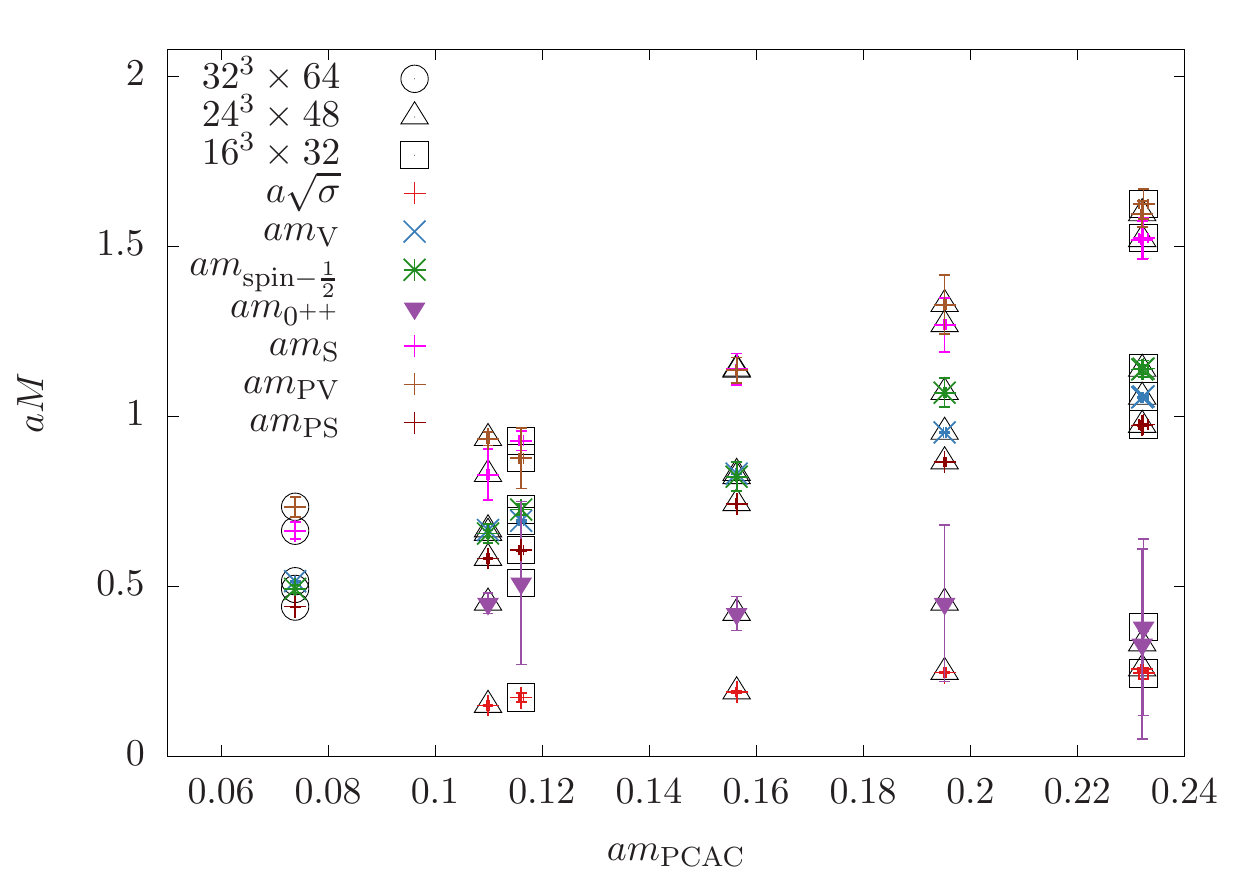}}
\subfigure{\includegraphics[width=0.49\textwidth]{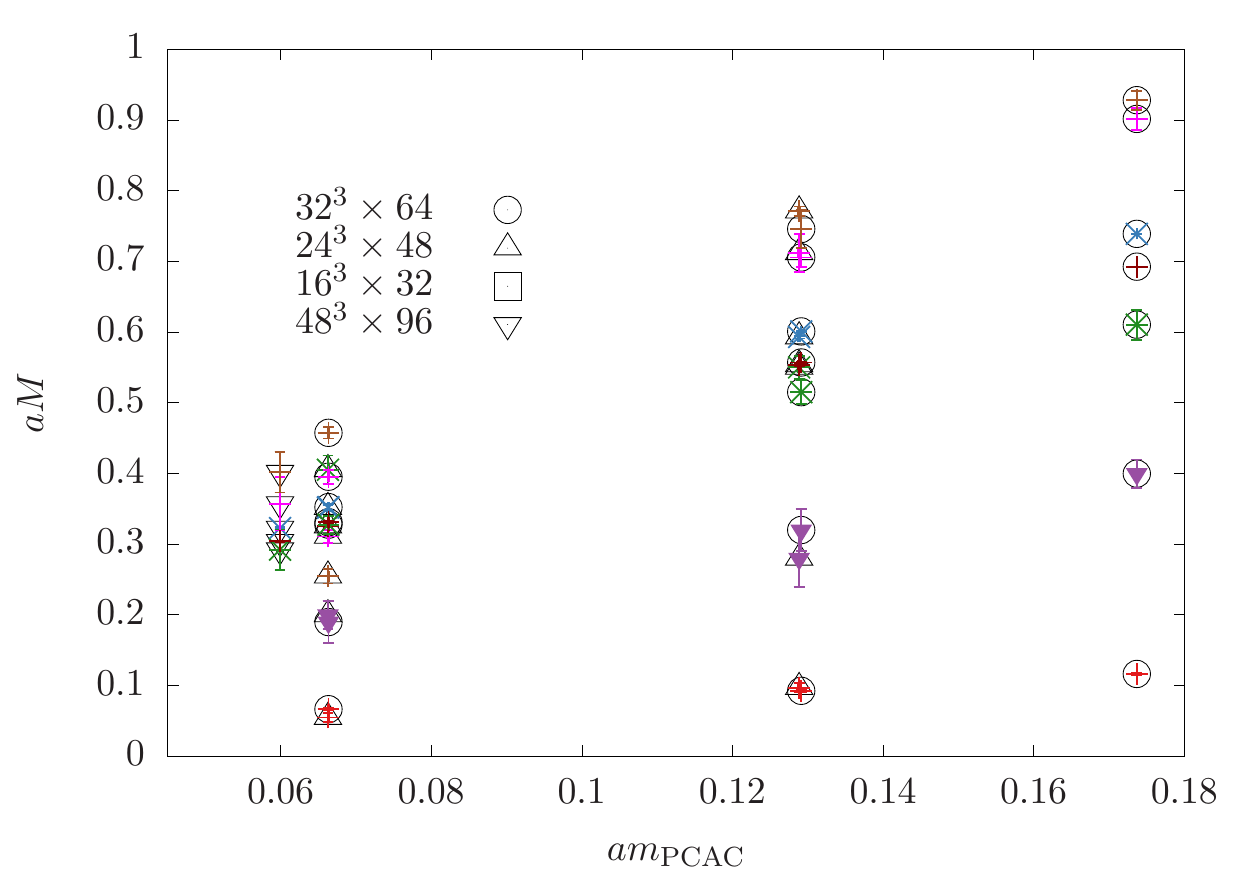}}
\caption{Particle masses and $\sqrt{\sigma}$ as a function of $a \mpcac$
for $\beta=1.5$ (left) and $\beta=1.7$ (right).}
\label{fig:masses}
\end{figure}
where $C$ is the charge conjugation matrix. Since one usually counts Dirac fermions, our fermions possess half the number of degrees of freedom as Dirac fermions and are counted as $N_f = 1/2$. Consequently $N_f = 3/2$ is to be interpreted as 3 species of Majorana fermions. In order to reduce lattice artifacts we use 3 levels of stout smearing \cite{Morningstar:2003gk} for the link fields in the Wilson-Dirac operator, with smearing parameter $\rho=0.12$.
For Majorana fermions the fermion integration
\begin{equation}
\int [d\psi]\ e^{- \frac{1}{2} \overline{\psi} D_w \psi}
= \mathrm{Pf}(C D_w) = \pm \sqrt{\det D_w}
\end{equation}
yields the Pfaffian of the Wilson-Dirac matrix. With 3 Majorana fermion fields the functional integrals contain a factor $(\det D_w)^{3/2}$, which
can be treated with the PHMC algorithm. The possible sign of Pf($C D_w$) has to be taken into account in the observables by reweighting. In simulations
not too close to the critical hopping parameter $\kappa_c$, negative signs are very rare and it was not necessary to consider them in the parameter
regions of our simulations for the determination of the masses.
For generating field configurations on the lattice we have used the two-step polynomial hybrid Monte Carlo (PHMC) algorithm \cite{Montvay:2005tj}.\newline
We have generated ensembles at two different values of the lattice coupling $\beta=1.5$ and $\beta=1.7$ and many different fermion masses. For a complete list of all used ensembles see \cite{Bergner:2017gzw}.
\section{Particle Spectrum and Scaling}
One way to determine if a theory is conformal is the dependence of the particle spectrum on the renormalized fermion mass $m_r$. In a theory outside the conformal window, with confinement and chiral symmetry breaking, the mass of the pseudo-Goldstone boson vanishes when the fermion mass $m_r$ goes to zero, whereas the other particle masses
approach a finite value.
In the IR conformal scenario all particle masses and the string tension would asymptotically scale to zero in the conformal limit
according to 
\begin{equation}
M \propto m_r^{1/(1 + \gamma^{\ast})}\,,
\end{equation}
where $\gamma^*$ is the value of the mass anomalous dimension at the fixed point \cite{Luty:2008vs,DelDebbio:2010ze}.
In this scenario the ratios of masses are approximately constant for small $m_r$. These ratios represent universal features of (near) IR conformal
theories \cite{Athenodorou:2016ndx}. In practice, however, the limit of vanishing fermion mass $m_r$ cannot be reached in numerical simulations. In a near conformal theory severe finite size effects would occur for small $m_r$, and have a substantial influence on the mass spectrum.
The part of the spectrum we investigated consists of scalar, pseudo-scalar, vector and pseudo-vector mesons, as well as the scalar glueball. In addition, we also find a spin $1/2$ fermion-glue bound state, represented by $\sigma_{\mu \nu} \text{tr}(F^{\mu \nu} \psi)$ possible due to the adjoint fermions. Apart from the particle masses we have also calculated the string tension $\sigma$ from the static quark-antiquark potential, where ``quark'' means a particle in the fundamental representation of the colour gauge group. The square-root of $\sigma$ has dimensions of a mass and scales as a mass.
\begin{figure}[t!]
\centering
\subfigure{\includegraphics[width=0.49\textwidth]{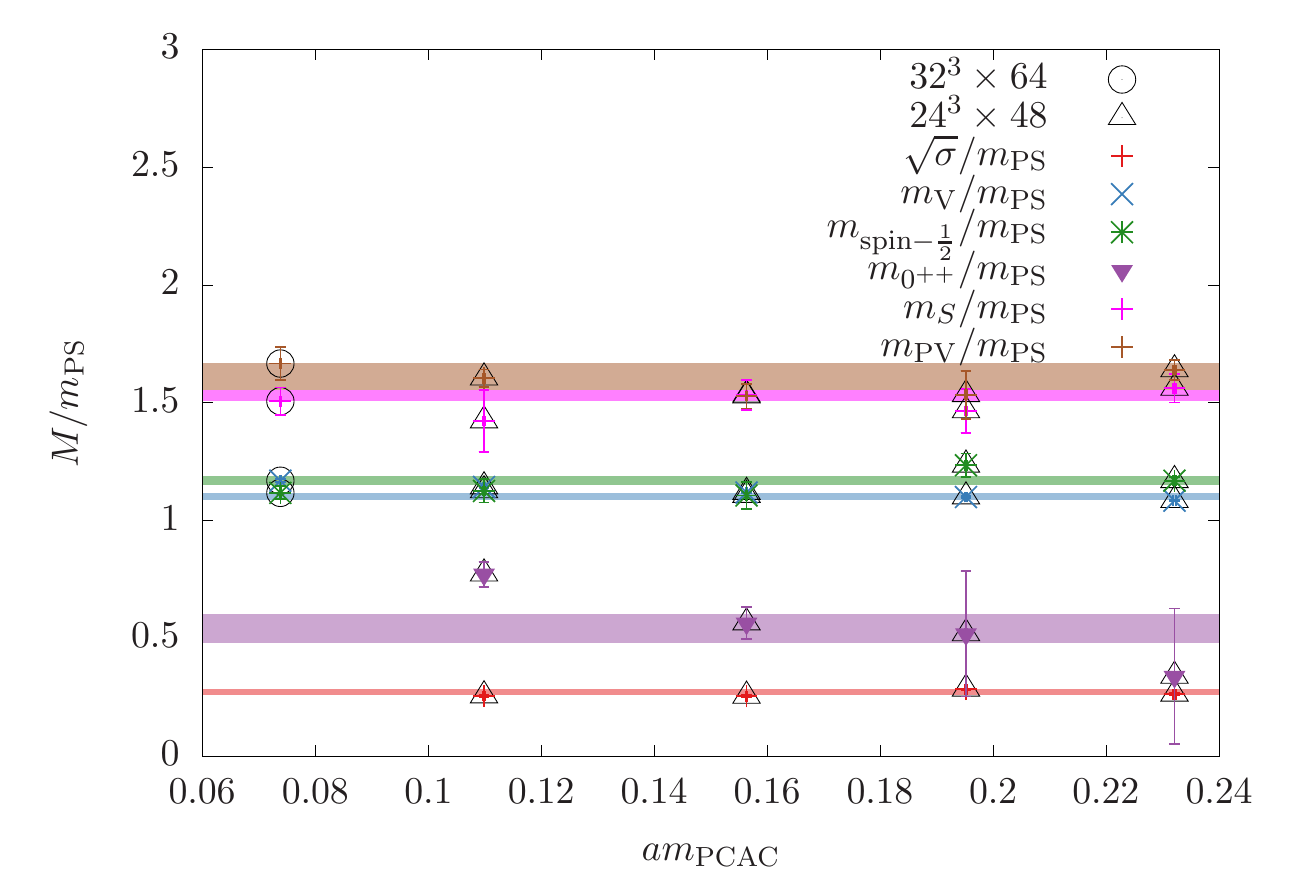}}
\subfigure{\includegraphics[width=0.49\textwidth]{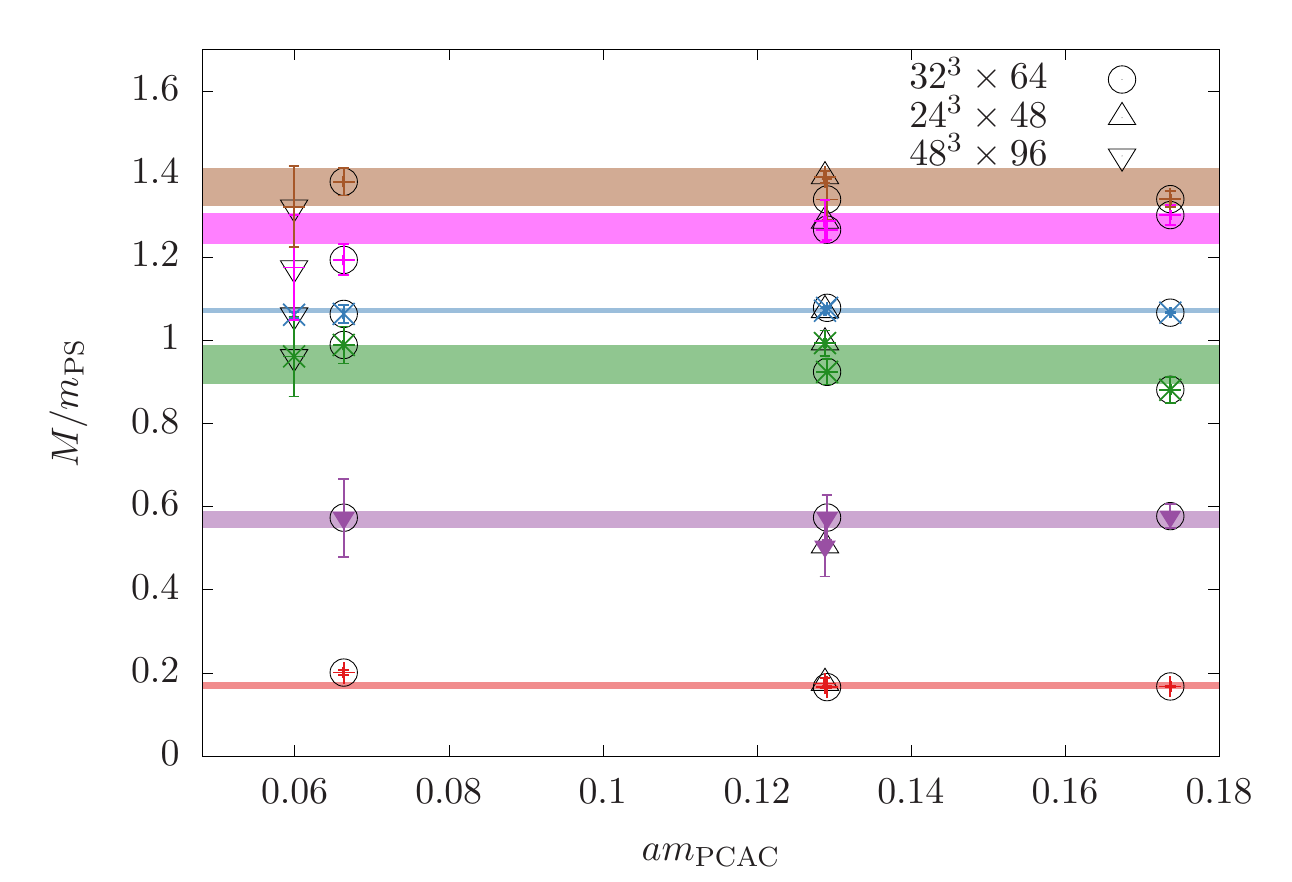}}
\caption{Particle masses and $\sqrt{\sigma}$ in units of the pseudoscalar
mass as a function of $a \mpcac$ for $\beta=1.5$ and $\beta=1.7$.}
\label{fig:ratios}
\end{figure}

Fig.~\ref{fig:masses} shows the particle masses as a function of the fermion mass. All masses appear to scale downwards towards the limit $\mpcac = 0$.
The lightest particle, being well separated from the rest, is the scalar glueball. The overall behaviour indicates a scenario different from the
QCD-like one, where the pseudo-scalar pseudo-Goldstone boson is lightest particle. As expected for a theory in the conformal window, all masses scale
approximately in the same way and their ratios are constant as shown in Fig.~\ref{fig:ratios}.
The estimate of the mass anomalous dimension can be obtained by a fit to the data. This gives $\gamma^{\ast}\approx 0.5$, for $\beta = 1.5$, and
$\gamma^{\ast} = 0.33(13)$, for $\beta = 1.7$.
\section{Mode number}
An alternative method for the determination of the mass anomalous dimension
is based on the spectral density of the Dirac operator
\cite{Giusti:2008vb,Patella:2012da,Cheng:2013bca,Fodor:2014zca,Cichy:2013eoa}.
The mode number $\nu(\Omega)$ is defined to be the number of eigenvalues of
the hermitian operator $D_w^\dag D_w$ below some limit $\Omega^2$. The mode
number obeys a scaling law \cite{Patella:2012da}
\begin{equation}
\label{eq:modenumber}
\nu(\Omega) = \nu_0 + a_1 (\Omega^2 - a_2^2)^{2/(1 + \gamma^{\ast})}
\end{equation}
for sufficiently small values of $\Omega^2 - a_2^2$, where $a_2$ is proportional to $\mpcac$. Therefore, a fit of $\nu(\Omega)$ to this function
in a suitable range $[\Omega_{\text{min}},\Omega_{\text{max}}]$ allows to estimate the mass anomalous dimension $\gamma^{\ast}$. 
The choice of the fit range $[\Omega_{\text{min}},\Omega_{\text{max}}]$ is a sensitive issue. For a small fit range near a scale $\Omega$, the resulting
value for the mass anomalous dimension can be considered as an effective anomalous dimension $\gamma(\Omega)$, which approximates the corresponding
renormalisation group function \cite{Cheng:2013bca}. For large $\Omega$ it is expected that $\gamma(\Omega)$ decreases and approaches its value zero at
the asymptotically free UV fixed point. On the other hand, for small $\Omega$ finite volume effects and effects of the non-vanishing fermion mass
$\mpcac$ will disturb the scaling behaviour. Therefore the fit range should be located in an intermediate regime, where the effects of the finite volume
and non-zero fermion mass can be neglected \cite{Patella:2012da,Cichy:2013eoa}. For an infrared conformal theory the coupling runs very slowly for a wide range of scales at low $\mu$, and there the anomalous dimension $\gamma$ varies slowly, too, approximatively developing a plateau at the value $\gamma^{\ast}$.
\begin{figure}[t!]
\centering
\subfigure{\includegraphics[width=0.49\textwidth]{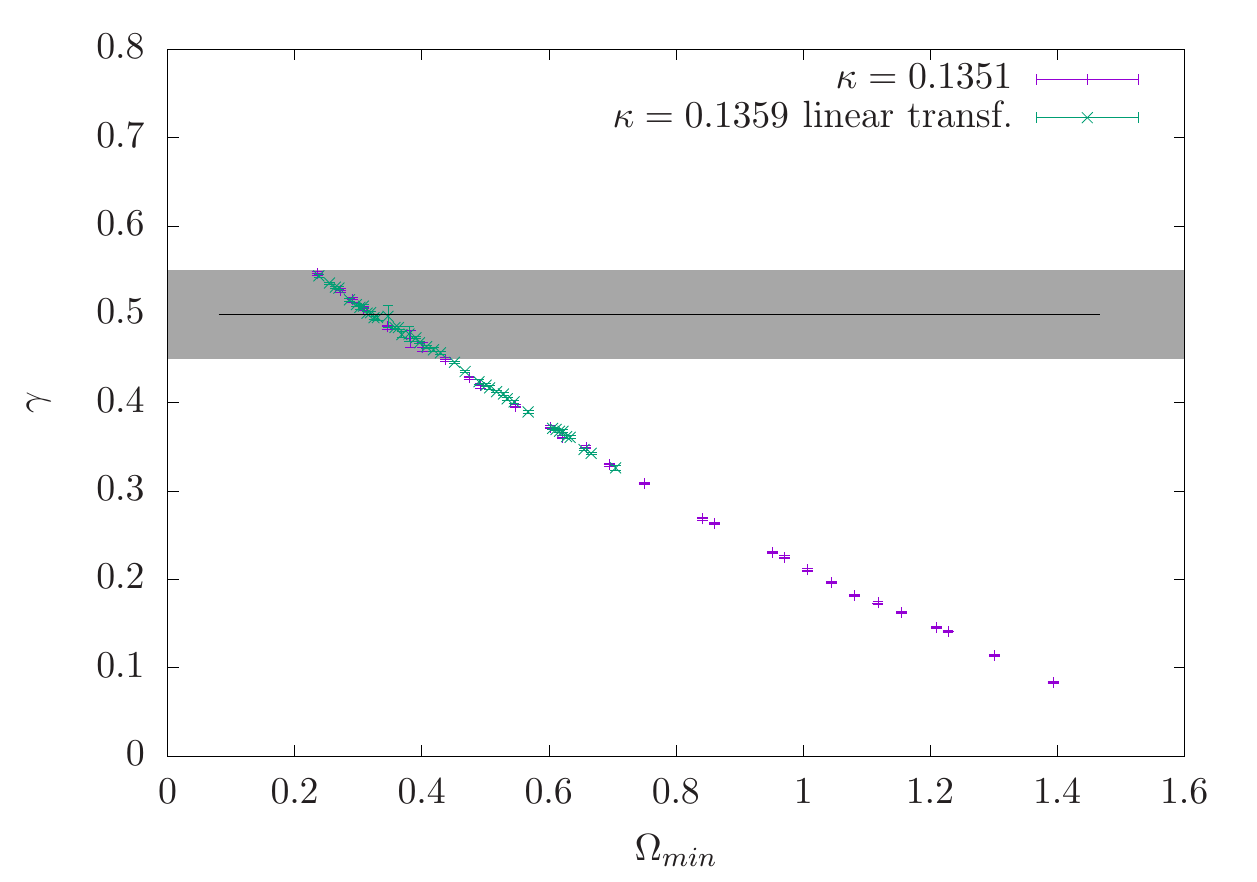}}
\subfigure{\includegraphics[width=0.49\textwidth]{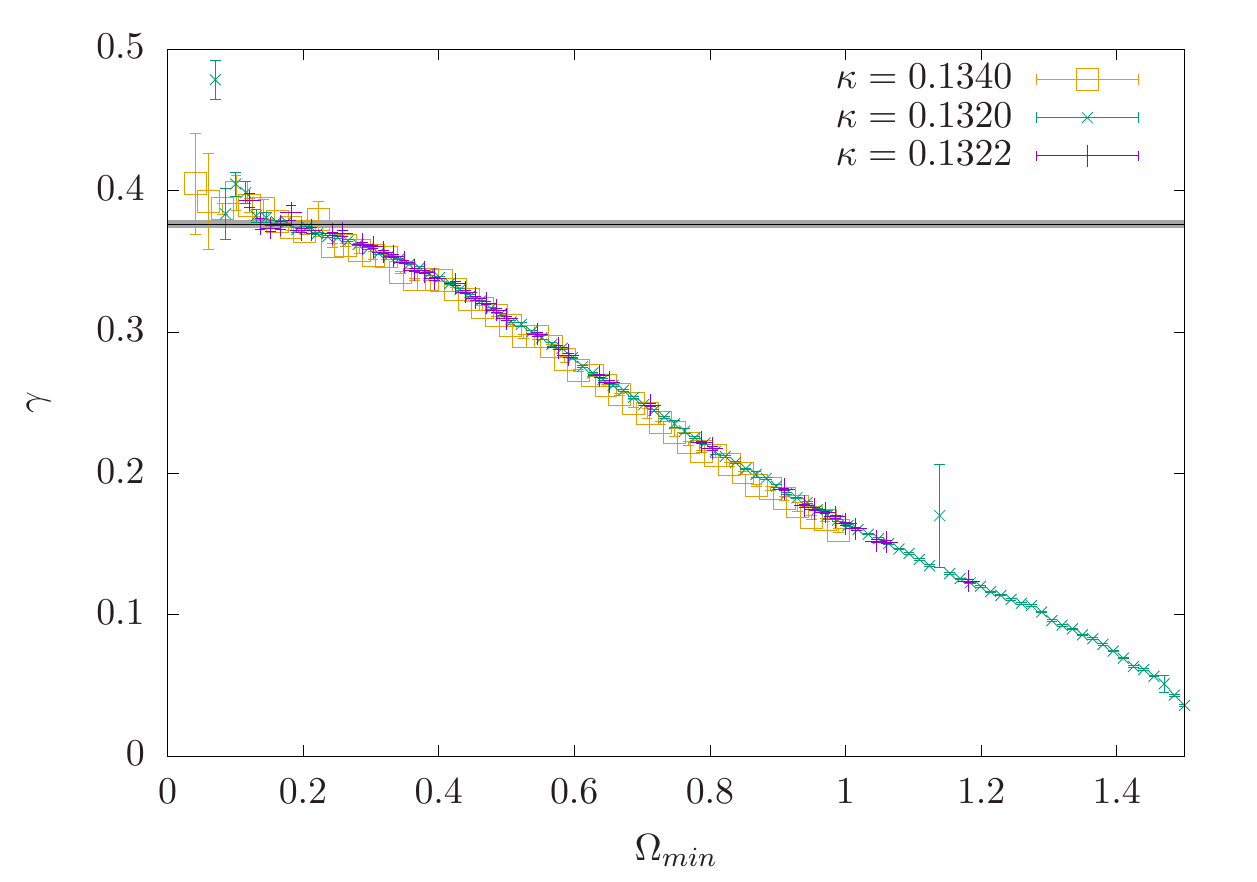}}
\caption{Fitted value of the mass anomalous dimension for the ensemble at
$\beta=1.5$, $\kappa=0.1351$ , and $\kappa=0.1359$.
$\Omega_{\text{min}}$ denotes the lower end of the fitting interval, while
the upper end is fixed to $\Omega_{\text{min}} + 0.07$. In the shaded region
we obtain the best fits for both ensembles (left). Fitted value of the mass anomalous dimension for the ensemble at
$\beta=1.7$, $\kappa=0.1322$, $\kappa=0.1320$, and
$\kappa=0.1340$ (right).}
\end{figure}
At $\beta=1.5$ we obtain reasonable fits with an acceptable p-value and a
correlated $\chi^2$ per degree of freedom in a certain region of $\Omega$
values for the ensembles with the smallest fermion masses. However, there is
no pronounced plateau for the obtained values in this range. The best fits
are obtained at rather large values of $\Omega$. Further in the infrared,
the correlated $\chi^2$ of the fit drastically increases, which is an
indication of fermion mass effects. We take the final value from the middle
of the range where the correlated $\chi^2$ per degree of freedom is below
2.5 for ensemble H, and the width of this range as an estimate for the
error. This provides a rough estimate of $\gamma^{\ast} \approx 0.5 \pm
0.05$.\newline
In contrast to the case of $\beta=1.5$, a considerable plateau of the fitted
values is obtained at $\beta=1.7$ in the infrared region. We obtain a value
of $\gamma^{\ast} \approx 0.377(3)$. Taking also the uncertainties in the
determination of the fitting interval into account, the estimate is
$\gamma^{\ast} \approx 0.38(2)$.

\begin{figure}[ht!]
\centering
\subfigure{\includegraphics[width=0.49\textwidth]{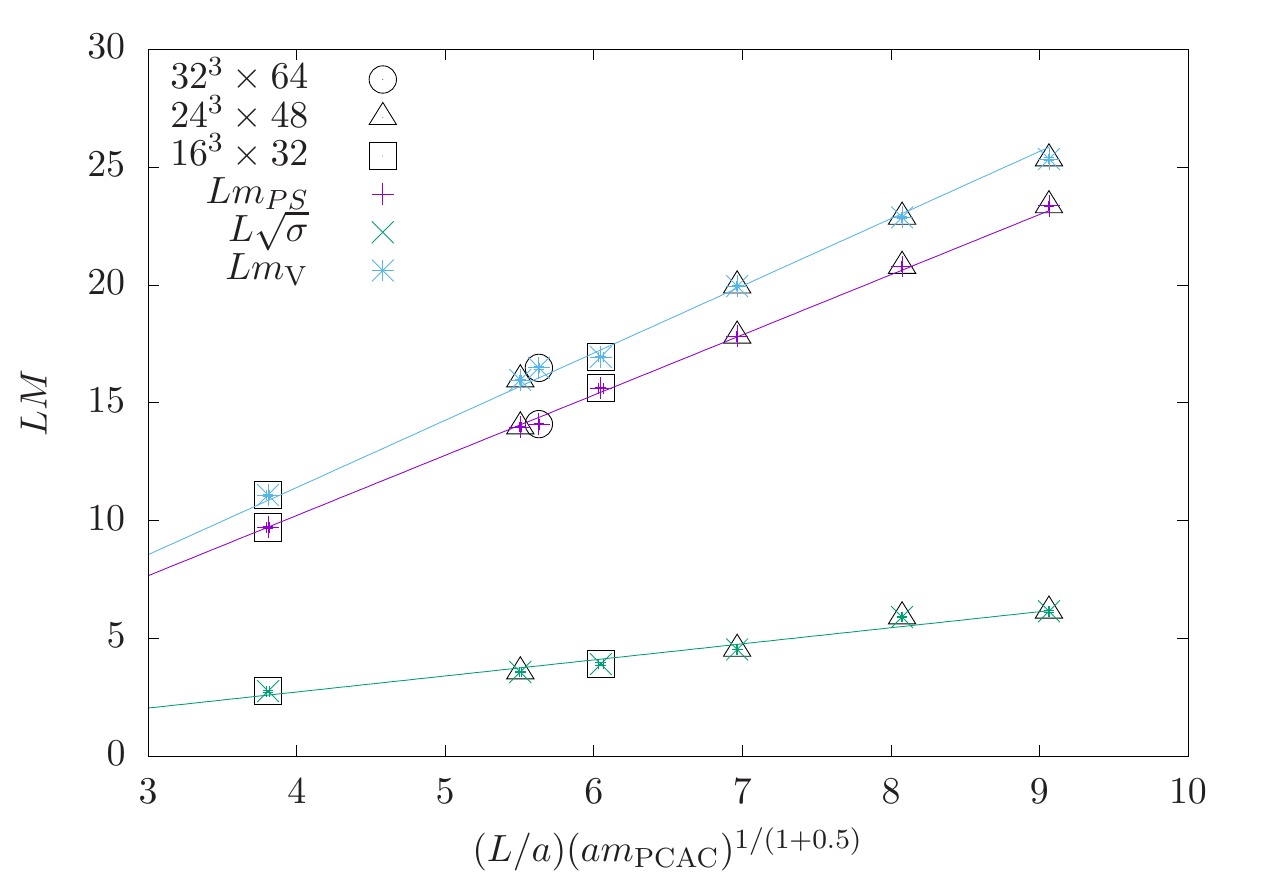}}
\subfigure{\includegraphics[width=0.49\textwidth]{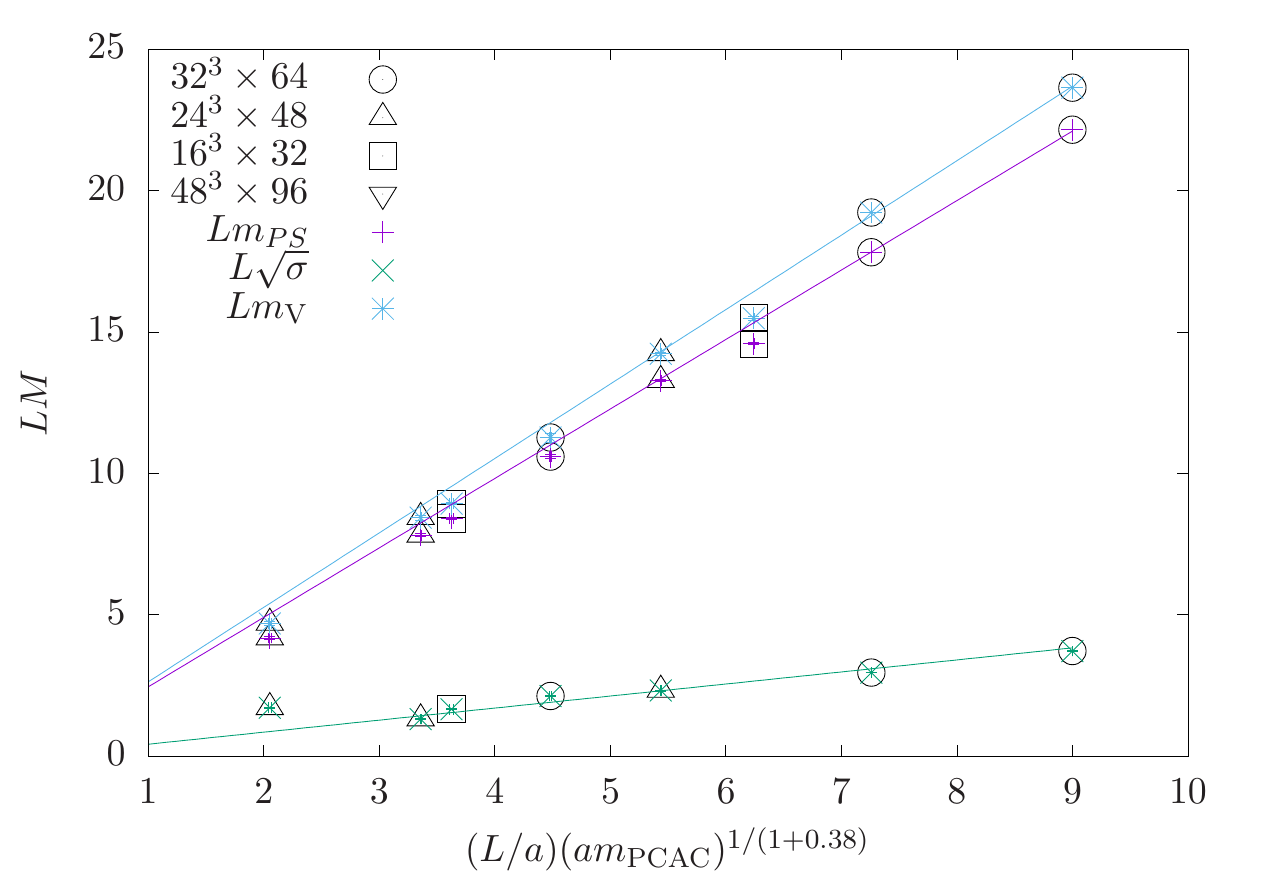}}
\caption{A cross check of the scaling exponents obtained from the mode
number with the scaling of the particle masses. These figures show a fit to
the hyperscaling hypothesis of the masses including volume scaling. The
points with the smallest values on the x-axis correspond to the ensembles B
($\beta=1.5$) and K ($\beta=1.7$). The lines correspond to a linear fit, in
case of $\beta=1.7$ without the data of ensemble K.}
\label{fig:scaling}
\end{figure}

We made a crosscheck of the obtained values of $\gamma^{\ast}$ with the
hyperscaling of the mass spectrum. As shown in Fig.~\ref{fig:scaling}, the
agreement with the expected functional behaviour is reasonable. We can also
vary the exponents close to the measured values in order to minimise the sum
of the $\chi^2$ from the linear fits. In this way we obtain a minimum around
$\gamma^{\ast} \approx 0.46(2)$ for $\beta=1.5$ and $\gamma^{\ast} \approx
0.37(2)$ for $\beta=1.7$. This shows that the values for the mass anomalous
dimension obtained from the mode number are consistent with the hyperscaling
of the mass spectrum.
\section{Running Coupling}
We can also try to look directly at the running coupling of the theory, to see 
whether the coupling runs, walks or does not change at all in some momentum range.
For this, we choose to fix our gauge to Landau gauge. As the running coupling in Landau
gauge is
\begin{equation}
    \alpha(p^2) = \alpha(\mu) Z(p^2) J^2(p^2) \, , \label{landau_coupl}
\end{equation}
where $Z(p)$ and $J(p)$ are the renormalized gluon and ghost dressing functions and $\mu$ is the
renormalization scale. We can simplify this even further by using the MiniMOM scheme \cite{Lorenz}, 
defined by $Z(\mu)=1, \, J(\mu)=1$. Then eq.~\eqref{landau_coupl} simplifies to
\begin{equation}
    \alpha(p^2) =\frac{g_\text{lat}}{4 \pi} Z_0(p^2) J_0^2(p^2) \, ,
\end{equation}
The running scale is defined by the lattice momentum and the renormalization scale is set at the
inverse lattice spacing. $Z_0(p^2)$ and $J_0(p^2)$ are the bare gluon and ghost dressing 
functions, that can be determined rather easily by measuring gluon and ghost propagators on
$O(20)$ gauge-fixed lattice configurations, well separated by more than fifty molecular dynamics units.
\begin{figure}[ht!]
\centering
\subfigure{\includegraphics[width=0.49\textwidth]{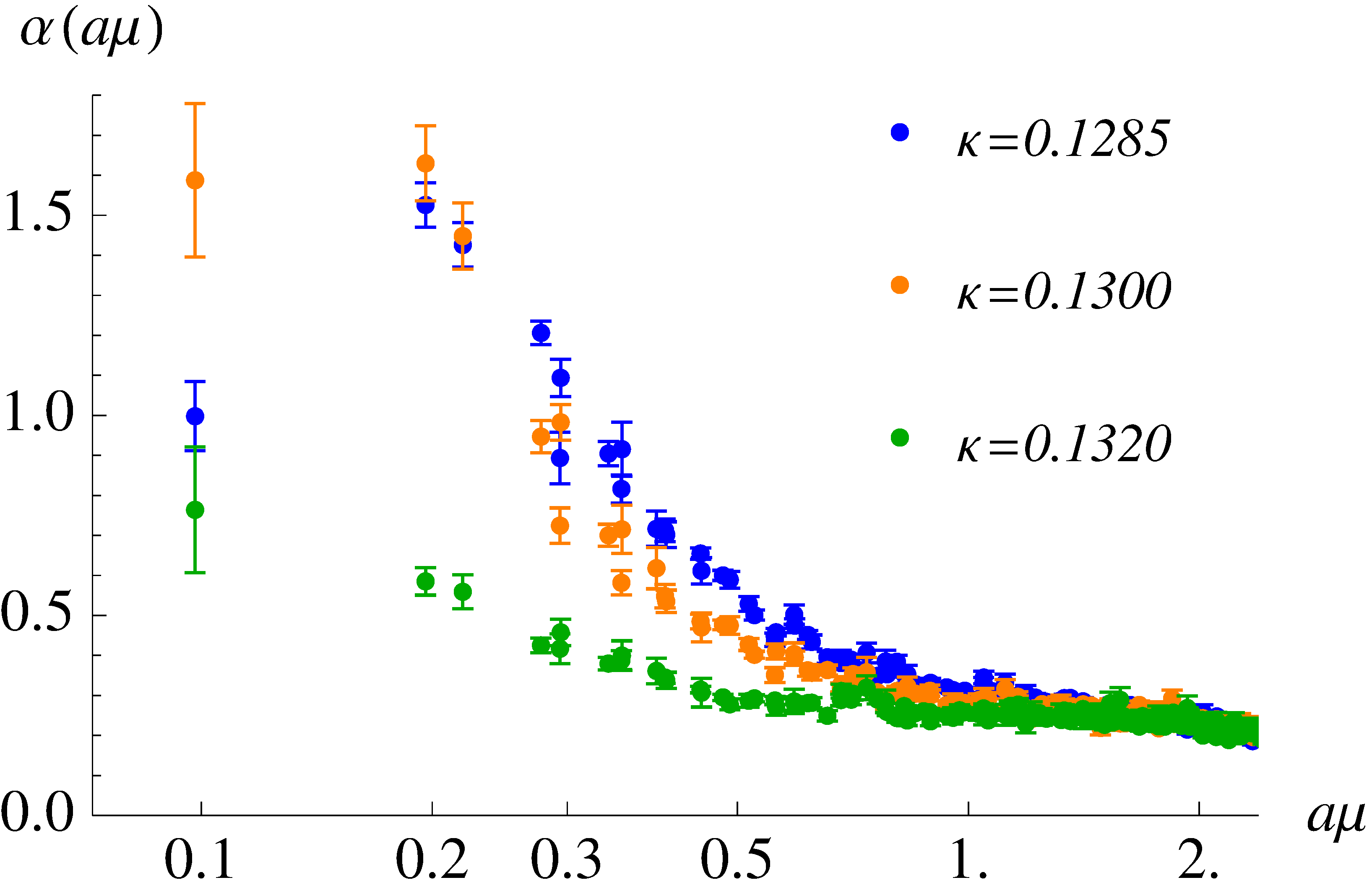}}
\subfigure{\includegraphics[width=0.49\textwidth]{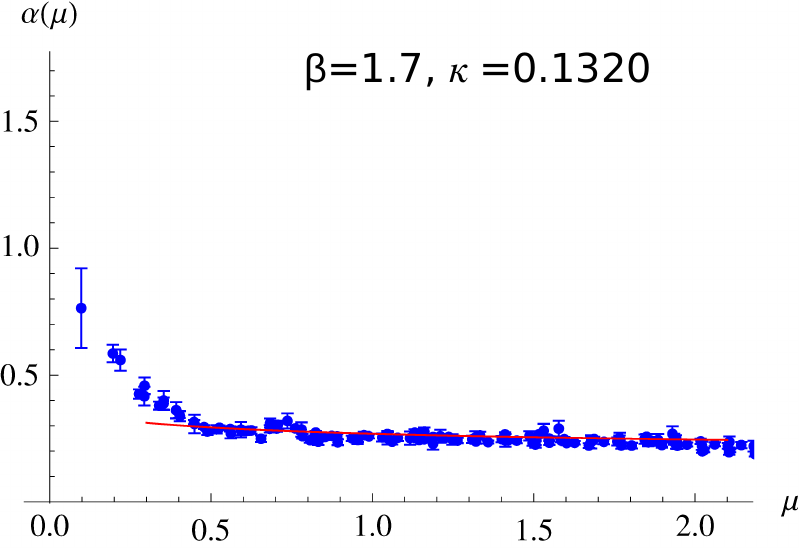}}
\caption{Running of the coupling for $\beta=1.7$ and several fermion masses (left). Fit of $\alpha(\mu)$ to 
perturbation theory for $\beta=1.7$ at the largest $\kappa$ (right).}
\label{fig:running_coupl}
\end{figure}
Fig.~\ref{fig:running_coupl} shows the running coupling for $\beta=1.7$ and three different values of $\kappa$.
For all three values of $\kappa$ the running coupling shows a peak and significant running in the infrared at this given lattice coupling.
We can also clearly see a dependence of the running coupling on the fermion mass. This is expected as the fermion mass
is a relevant parameter of the theory that actually pushes the theory away from the conformal fixed point. Also in agreement with this fact,
the running of the coupling decreases with the fermion mass and at higher momentum where the effects of the non-vanishing fermion mass can be neglected.
Even though the running of the coupling is slow in the large momentum region it is still incompatible with zero.
The running of the coupling can also be fitted to perturbation theory \cite{Bergner:2017ytp, Rytov} to check a possible relation to the continuum theory. Doing this we can extract the
$\Lambda$-parameter of the theory. We find
\begin{align*}
a \Lambda^\text{MiniMOM}(\beta_\text{lat}=1.5) &= 0.0051(8) \; , \\
a \Lambda^\text{MiniMOM}(\beta_\text{lat}=1.7) &= 0.0054(5) \; . 
\end{align*}
All quoted errors are purely systematic corresponding to variations coming from different
choices of the fitting intervals. The value of $\Lambda$ in lattice units is significantly smaller than
the masses of hadrons in lattice units at the same parameters, for instance the glueball $0^{++}$ mass $am_{0^{++}} \approx 0.25(3)$.
The contrast to the well-known confining gauge theories is quite significant, in QCD the value of the intrinsic ultraviolet scale $\Lambda$
is comparable to the hadron scales in the various MOM schemes. It is, furthermore, remarkable how weak the dependence of $a \Lambda$
on the bare lattice coupling is. Even though these observations are not enough to provide a strong evidence for a conformal behavior, they are clearly much different from QCD-like theories.

\end{document}